\documentclass[twocolumn]{bmcart}
\usepackage{booktabs}
\newcommand{\tabincell}[2]{\begin{tabular}{@{}#1@{}}#2\end{tabular}}
\usepackage{graphicx}
\usepackage{algorithm}  
\usepackage{algorithmic}
\usepackage{ragged2e}

\usepackage{amsthm,amsmath}

\usepackage[utf8]{inputenc}

\startlocaldefs
\endlocaldefs

\begin{document}

\begin{frontmatter}

\begin{fmbox}
\dochead{Research}

\author[
  addressref={aff1},
  corref={aff1},
  email={lxinrui10@gmail.com}
]{\inits{X.R.L.}\fnm{Xinrui} \snm{Liu}}
\author[
  addressref={aff1},
  email={wangyajie0312@foxmail.com}
]{\inits{Y.J.W.}\fnm{Yajie} \snm{Wang}}
\author[
  addressref={aff1},
  email={tan2008@bit.edu.cn}
]{\inits{Y.A.T.}\fnm{Yu-an} \snm{Tan}}
\author[
  addressref={aff1},
  email={kfqiu@bit.edu.cn}
]{\inits{K.F.Q.}\fnm{Kefan} \snm{Qiu}}
\author[
  addressref={aff1},
  email={popular@bit.edu.cn}
]{\inits{Y.Z.L.}\fnm{Yuanzhang} \snm{Li}}

\address[id=aff1]{%
  \orgdiv{School of Cyberspace Science and Technology},
  \orgname{Beijing Institute of Technology},          
  \city{Beijing},                              
  \cny{China} 
}

\title{Towards Invisible Backdoor Attacks in the Frequency Domain against Deep Neural Networks}

\begin{abstractbox}

\begin{abstract} 
\justifying
Deep neural networks (DNNs) have made tremendous progress in the past ten years and have been applied in various critical applications. However, recent studies have shown that deep neural networks are vulnerable to backdoor attacks. By injecting malicious data into the training set, an adversary can plant the backdoor into the original model. The backdoor can remain hidden indefinitely until activated by a sample with a specific trigger, which is hugely concealed, bringing serious security risks to critical applications. However, one main limitation of current backdoor attacks is that the trigger is often visible to human perception. Therefore, it is crucial to study the stealthiness of backdoor triggers. In this paper, we propose a novel frequency-domain backdooring technique. In particular, our method aims to add a backdoor trigger in the frequency domain of original images via Discrete Fourier Transform, thus hidding the trigger. We evaluate our method on three benchmark datasets: MNIST, CIFAR-10 and Imagenette. Our experiments show that we can simultaneously fool human inspection and DNN models. We further apply two image similarity evaluation metrics to illustrate that our method adds the most subtle perturbation without compromising attack success rate and clean sample accuracy.
\end{abstract}


\end{abstractbox}
\end{fmbox}

\end{frontmatter}

\section*{Introduction}
\begin{figure}
  \includegraphics[width=0.9\linewidth]{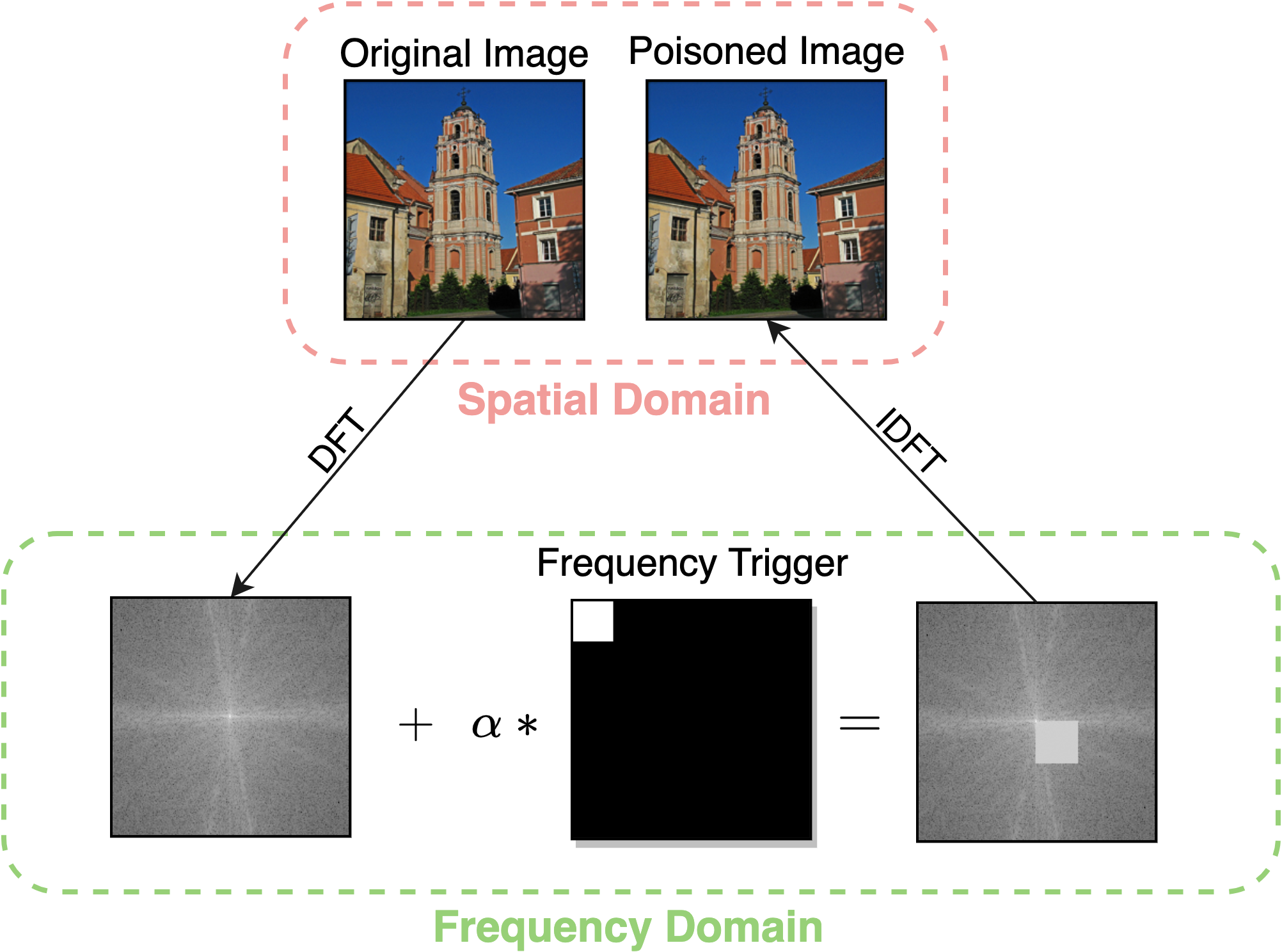}
  \caption{Overview of our method. In the figure, DFT and IDFT represents Discrete Fourier Transform and Inverse Discrete Fourier Transform respectively. Note that we shift the zero-frequency component to the center of the spectrum.}%
  \label{figs:DFT_generator_overview}
\end{figure}

With the advent of artificial intelligence, neural networks have become a widely used method of artificial intelligence. Currently, neural networks have been adopted in a wide range of areas, such as face recognition~\cite{russakovsky2015imagenet}, voice recognition~\cite{graves2013speech}, games~\cite{hermann2013multilingual}, and autonomous driving~\cite{bojarski2016end}. For example, PayPal users are using deep learning-based facial recognition systems to make payments. However, recent studies have shown that deep learning models are vulnerable to various attacks. Attacks against DNN~\cite{schmidhuber2015deep} can be divided into three classes: adversarial example, poisoning attack, and backdoor attack. Adding some perturbation to the input data, an adversarial attack~\cite{szegedy2013intriguing} can cause misclassification by the DNN without affecting the DNN. However, this attack generates perturbations specific to a single input. Poisoning attack~\cite{biggio2012poisoning} is a method that reduces the accuracy of the model by injecting malicious training data during the training phase. However, this method only reduces the accuracy of the model. Attackers cannot choose specific data they want to cause misclassification. Also, users will not deploy models with low accuracy under normal circumstances, which brings limitations in practice. To overcome these problems, backdoor attack~\cite{li2020backdoor} is proposed. \par
The backdoor attack enables attackers to plant a backdoor into the model and performs malicious attacks using a specific backdoor trigger in the inference phase. The backdoored deep neural network can correctly classify benign samples but will misclassify any input with a specific backdoor trigger as an attacker chosen target. The backdoor can remain hidden indefinitely until activated by a sample with a specific backdoor trigger, which is hugely concealed. Therefore, it can bring serious security risks to many critical applications.\par
Although backdoor attacks have been proven to cause neural network misclassifications successfully, one main limitation of current backdoor attacks is that backdoor triggers are usually visible to human perception. When the system administrator manually checks these datasets, the poisoned data will be found suspicious. ~\cite{chen2017targeted} first discussed the importance of improving the stealthiness of backdoor triggers. They designed a method to blend the backdoor trigger with benign inputs instead of stamping the trigger as proposed in conventional backdoor attack~\cite{gu2017badnets}~\cite{liu2017trojaning}. After that, there was a series of researches dedicated to the invisibility in the backdoor attack. However, the backdoor inputs are still noticeable compared to benign samples, making existing backdoor triggers less feasible in practice. Therefore, improving the invisibility of backdoor triggers has become a research hotspot of neural network backdoor attacks. The challenge of creating an invisible backdoor is how to achieve smaller perturbation without affecting the attack success rate and clean sample accuracy.  In 2019, ~\cite{saha2020hidden} exploit the backdoor attack in a robust manner, namely hidden trigger backdoor. Here, the trigger is invisible to evade human inspections. However, we perform several experiments to prove that the perturbations they add are relatively large in contrast to our method. Besides, the adversary utilizes a neural network to optimize the original samples to generate poisoned samples, which raises the attack cost compared to our method.\par

It is well known that humans cannot perceive subtle variations in the color space within images. However, deep neural networks can detect slight perturbation due to their complexity and powerful feature extraction capabilities, making it possible to hide the trigger from manual review. Therefore, in this paper, we exploit this characteristic of DNNs to implement invisible backdoor attacks. Our method is motivated by the DFT-based image blind watermark. In this technique, a sender hides the covert information in the image frequency domain using an encoder. A receiver applies a decoder to extract the hidden message from the frequency domain to achieve secret messaging. According to our investigations, we are the first to propose the frequency-domain backdooring techniques. Figure~\ref{figs:DFT_generator_overview} demonstrates an overview of our method. We add a backdoor trigger in the frequency domain of an original image to generate a poisoned sample which is invisible enough to evade human perception.\par  
Our experimental results show that we can simultaneously achieve invisible backdoor attack without affecting attack success rate and clean sample accuracy. Also, we apply two image similarity evaluation metrics ($l_{2}$ paradigm and LPIPS (Learned Perceptual Image Patch Similarity)~\cite{zhang2018unreasonable}) to compare our method with the conventional method and a state-of-the-art hidden trigger attack~\cite{saha2020hidden}. We found that our method adds the smallest perturbation without compromising attack performance.\par
The contributions of this paper are as follows:\par

$\bullet$ We propose the first class of frequency-domain backdooring techniques in which our method aims to add a backdoor trigger in the frequency domain of original images via Discrete Fourier Transform (DFT), thus hidding the trigger.\par
$\bullet$ We implement our DFT-based backdoor attack on MNIST, CIFAR-10, and a subset in Imagenet. Our experimental results show that our approach can simultaneously fool human inspection and DNN models.\par
$\bullet$ We apply two image similarity evaluation metrics ($l_{2}$ paradigm and LPIPS) to compare the invisibility of different methods. We find that our method adds the smallest perturbation without sacrificing attack success rate and clean sample accuracy.\par

The rest of the paper is organized as follows. Section 2 describes the related work. Section 3 explains the proposed scheme. Section 4 demonstrates the experimental setup and evaluates the results. Finally, we conclude the paper in Section 5.

\begin{figure*}[tbp]
    \centering
    \includegraphics[width=2\linewidth]{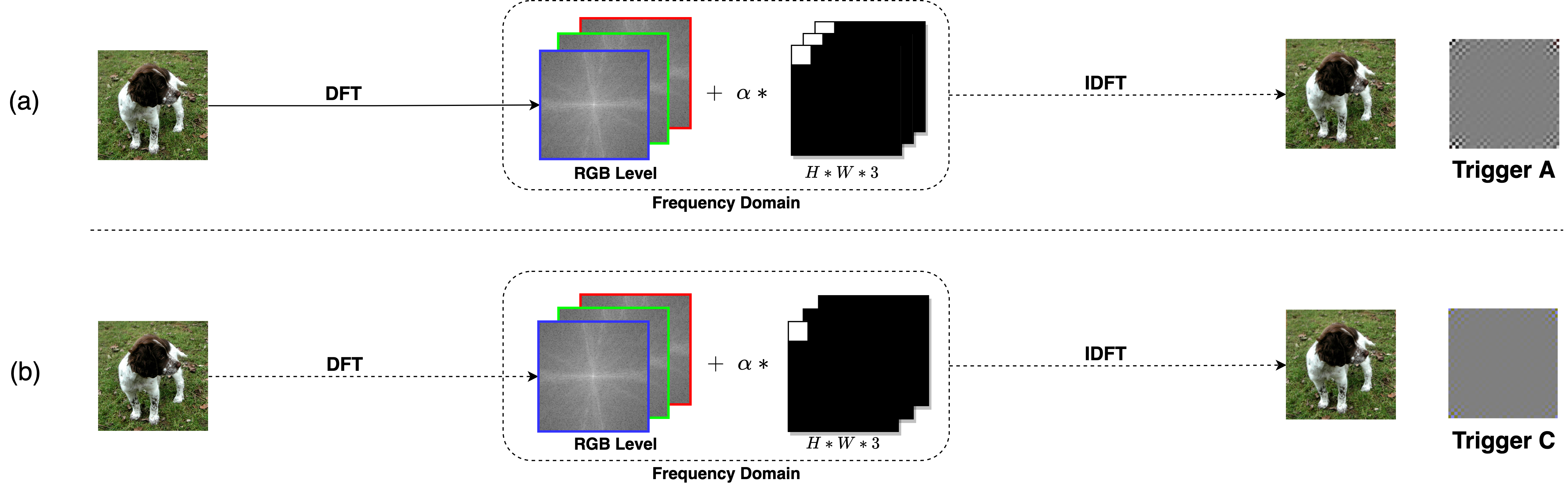}
    \caption{Two different DFT-based methods of generating poisoned samples for RGB images. The generation of Trigger B applies a RGB-to-Gray transformation, thus further improving the invisibility.}%
    \label{fig:DFT_RGB}
\end{figure*}

\section*{Related Work}
\subsection*{Backdoor attack against DNNs}
Backdoor attacks were first migrated to neural networks in 2017. ~\cite{gu2017badnets} proposed BadNets. In this method, the attacker can attach a specific trigger to the stop sign image and mark it as the speed limit sign to generate a backdoor in the road sign recognition model. Although the model can correctly classify clean samples, it will misclassify the stop sign image with the trigger as the speed limit.\par
In 2018, ~\cite{liu2017trojaning} proposed a more advanced backdoor attack technique called Trojan attack. In the study of the Trojan attack, it was found that the backdoor attack method in the neural network was effective because the backdoor trigger would activate specific neurons in the network. Therefore, the Trojan attack generates a trigger in a way that maximizes the activations of specified neurons. \par
Based on classical backdoor attacks, many works focused on improving the invisibility of backdoor images. ~\cite{chen2017targeted} first discuss the importance of invisibility in backdoor attacks. They proposed that a backdoored image should be indistinguishable from its benign version to evade human inspection. To satisfy such a requirement, they generated poisoned images by blending the backdoor trigger with benign inputs rather than stamp the trigger as proposed in conventional backdoor attacks. After that, there was a series of researches dedicated to the invisibility in backdoor attacks. ~\cite{turner2019label} proposed to utilize a backdoor trigger amplitude to perturb the clean images instead of replacing the corresponding pixels with the chosen pattern.\par
Interestingly, ~\cite{saha2020hidden} exploit the backdoor attack in a robust manner, namely, hidden trigger backdoor. In this method, the trigger used in the poisoning phase is invisible to evade human inspections. However, we perform several experiments to prove that the perturbations they add are relatively large in contrast to our method, making it easily detected by programs. Besides, the attacker utilizes a neural network to optimize the original samples to add perturbations, which raises the attack cost to generate poisoned samples compared to our method.\par
In order to evaluate the invisibility of our method, we investigate a series of methods used to calculate image similarity, such as phash, $l_{2}$ paradigm, $l_{\infty}$ paradigm, and so on. Among them, LPIPS~\cite{zhang2018unreasonable} is used to measure the similarity between two images in a manner that simulates human judgment. LPIPS is proposed based on perceptual loss. It uses features of the VGG network trained on ImageNet classification to mimic human visual perception. In this paper, we will use LPIPS as an invisibility evaluation metric.\par
\subsection*{Blind Watermark}
Blind watermark is an algorithm in steganography~\cite{cox2007digital} which is the study of concealing information in plain sight, such that only the intended recipient would get to see it. Steganography encodes hidden messages onto conventional multimedia data, which may be an image, text, and video. One widely used algorithm in steganography is the Least Significant Bit (LSB) substitution. The idea behind LSB is that replacing bit 0 (i.e., the lowest bit) in a binary pixel value will not cause a visible change in the color space. Though this spatial-domain technique has the least complexity and high payload, it cannot withstand image compression and other typical image processing attacks, which bring poor robustness.\par
The frequency-domain blind watermark based on the Discrete Fourier Transform (DFT)~\cite{pun2006novel} typically provides imperceptibility and is much more robust to image manipulations. The DFT-based blind watermark's main idea is to add a watermark image in the original image's frequency domain using DFT and transform the frequency-domain image back to spatial-domain using Inverse Discrete Fourier Transform (IDFT). Note that the frequency-domain image demonstrates the intensity of image transformation.

\section*{Methodology}
\subsection*{Threat model}
We assume a user who wants to use a training dataset $D_{train}$ to train the parameters of a DNN. The user sends the internal structure of the DNN $M$ to the trainer. Finally, the trainer will return to the user the trained model parameters $\Theta^{'}$.\par
However, the user cannot fully trust the trainer. The user needs to check the accuracy of the trained model on the validation dataset $D_{valid}$. Only when the model's accuracy meets an expected accuracy rate $r^*$ will the user accept the model.\\
\textbf{Attacker's Goals:} The attacker expects to return to the user a maliciously trained backdoor model parameters $\Theta^{'}:=\Theta^{adv}$. The parameters of this model are different from those of the honestly trained model. A backdoored model needs to meet two goals:\par
Firstly, the classification accuracy of the backdoored model $M_{{\Theta^{adv}}}$ cannot be reduced on the validation set $D_{valid}$, in other words, that is, $C(M_{\Theta^{adv}},D_{valid})\ge r^*$. Note that the attacker cannot directly access the user's validation dataset.\par
Secondly, for the input containing the backdoor trigger specified by the attacker, $M_{\Theta^{adv}}$ outputs' predictions are different from the outputs of the honestly trained model.\par
\subsection*{Generate poisoned images with DFT}
In conventional backdoor trigger design approaches, the backdoor trigger is usually a distinct sign within an area, making backdoor data easily recognizable in the event of a human visual inspection. Our approach is inspired by the DFT-based image blind watermark~\cite{eggers2001blind} in image steganography~\cite{cox2007digital}. Similarly, we add a trigger to an image's frequency domain so that the perturbation spreads throughout the image instead of being confined to a fixed region, thus making the trigger more invisible. \par
\begin{align}
    \begin{split}
    F(u,v)&=DFT(f(p,q))\\&=\sum_{p=0}^{H-1}\sum_{q=0}^{W-1}f(p,q)e^{-i2\pi(\frac{up}{H}+\frac{vq}{W})}
    \end{split}
    \label{Equation:DFT}
\end{align}
\begin{align}
    \begin{split}
    f(p,q)&=IDFT(F(u,v))\\&=\frac{1}{HW}\sum_{u=0}^{H-1}\sum_{v=0}^{W-1}F(u,v)e^{i2\pi(\frac{up}{H}+\frac{vq}{W})}
    \end{split}
    \label{Equation:IDFT}
\end{align}\par

We assume that we have a grayscale image that can be viewed as an $H*W$ matrix ($H$, $W$ denote the height and width of the image, respectively). We can regard this image as a signal $f(p,q)$ (denotes the pixel value of the spatial domain image at the coordinate point $(p,q)$). In digital image processing, we usually utilize Discrete Fourier Transform (DFT) to convert an image from spatial domain to frequency domain. Besides, we apply $F(u,v)$ to denote the pixel value of an image in frequency domain at the coordinate point $(u,v)$. The following Equation~\ref{Equation:DFT} represents Discrete Fourier Transform, and Equation~\ref{Equation:IDFT} represents the Inverse Discrete Fourier Transform (IDFT), which transforms an image from frequency domain to spatial domain. Note that $i$ denotes a unit of the complex number. \par
As shown in Algorithm~\ref{algorithm:DFT-based_backdoor_attack}: line 4 to line 8, we define a trigger $F_{trigger}$ in frequency domain and the original image in spatial domain is represented as $f_{original}$. We first convert the original image $f_{original}$ to frequency domain using DFT (Equation~\ref{Equation:DFT}), the result is represented as $F_{original}$. Then, we add a trigger in the frequency-domain image $F_{original}$ to generate a poisoned image of its frequency form. Here, we define an energy factor $\alpha$ to indicate the strength of the trigger. The smaller the $\alpha$, the lower the visibility of the trigger. Finally, we convert the poisoned image in frequency domain back to spatial domain by performing IDFT (Equation~\ref{Equation:IDFT}). $f_{poisoned}$ is our generated spatial-domain backdoor image. Figure~\ref{figs:DFT_generator_overview} demonstrates the visualization of our algorithm.\par
For RGB images, we design two approaches to add triggers in the frequency domain. One is to add the trigger directly in RGB-level frequency domain of the original image; the shape of the trigger is $H*W*3$. The added perturbation is shown in Figure ~\ref{fig:DFT_RGB}(a). In the second method, we first convert the RGB image to grayscale and then add a trigger (Note that the trigger shape here is $N*M$) in the grayscale frequency domain. Finally, we convert the gray image back to RGB-level, as shown in Figure~\ref{fig:DFT_RGB}(b). \par
\subsection*{Backdoor injection}
After generating DFT-based poisoned images, as shown in Algorithm~\ref{alg:algorithm}: line 9, we replace the labels of the poison samples generated in Section 3.2 with the target label $t$. After that,  we can obtain a poisoned dataset $D_{poisoned}$. We apply the poisoned dataset $D_{poisoned}$ with the clean dataset $D_{clean}$ to retrain the model parameters $\Theta^{adv}:=\Theta^{'}$.\par
In the inference phase, we apply the same frequency-domain trigger and $\alpha$ value used in the training phase to generate poisoned validation samples. After that, we record the Clean Sample Accuracy (CSA) as well as the Attack Success Rate (ASR) to evaluate our attack. We will show our experiment results in the next section.
\begin{algorithm}[tb]
\caption{DFT-based Backdoor attack}
\label{alg:algorithm}
\textbf{Input}: Frequency trigger: $F_{trigger}$, Original model's internal structure: $M$, Original training images: $X$ and its corresponding label set: $Y$, Original training set: $D_{train}=(X,Y)$, Attack target: $t$\\
\textbf{Parameter}: Energy factor: $\alpha$, Pollution rate: $\beta$\\
\textbf{Output}: Retrained model's parameter: $\Theta^{adv}$\\
\begin{algorithmic}[1] 
\STATE Select $\beta*D_{train}$ as poisoned dataset $D_{poisoned}$ and $(1-\beta)*D_{train}$ as clean dataset $D_{clean}$.
\FOR{$(x_{i},y_{i})$ in $D_{poisoned}$}
    \STATE $f_{original}:=x_{i}$
    \STATE Transform $f_{original}$ to frequency domain using DFT (Equation~\ref{Equation:DFT}).\\
    $F_{original}:=DFT(f_{original})$
    \STATE Add $F_{trigger}$ to $F_{original}$ and use $\alpha$ to control the trigger visibility.\\
    $F_{poisoned}:=F_{original}+\alpha*F_{trigger}$
    \STATE Transform $F_{poisoned}$ to spatial domain using IDFT (Equation~\ref{Equation:IDFT}).\\
    $f_{poisoned}:=IDFT(F_{poisoned})$\\
    \STATE Normalize\ $f_{poisoned}$ to\ $[0,1.0]$\\
    \STATE $x_{i}=f_{poisoned}$\\
    \STATE $y_{i}=t$\\
\ENDFOR
\STATE Retrain target classifier parameter.\\
    $\Theta^{adv}\leftarrow D_{poisoned}+D_{clean}$\\
\STATE \textbf{return} $\Theta^{adv}$
\label{algorithm:DFT-based_backdoor_attack}
\end{algorithmic}
\label{algorithm:DFT-based_backdoor_attack}
\end{algorithm}

\section*{Experiments and Analysis}

\subsection*{Experiment setup}
In this section, we implement the DFT-based backdoor attack introduced in Section 3. 
\paragraph{Datasets and models.}For the DFT-based backdoor attack, we mount our attack on MNIST~\cite{lecun1998mnist}, CIFAR-10~\cite{krizhevsky2009learning}, and Imagenette which is a subset in ImageNet~\cite{deng2009imagenet}. All datasets are widely used in deep learning. Our experiments were run on a machine with two 2080Ti, and our networks are implemented by Pytorch 1.5~\cite{paszke2019pytorch}.
For MNIST digit recognition task, in order to obtain high classification accuracy, we use AlexNet~\cite{krizhevsky2017imagenet} as our baseline model. For CIFAR-10 and Imagenette, we use pre-trained ResNet-18~\cite{he2016deep} as the original model. Note that we use Adam~\cite{kingma2014adam} on Alexnet with a learning rate of $1e-3$ and apply SGD~\cite{ruder2016overview} optimizer on ResNet-18 with a learning rate of $1e-2$.
\paragraph{Evaluation metric.}The success of a backdoor attack can be generally evaluated by Clean Sample Accuracy(CSA) and Attack Success Rate(ASR), which can be defined as follows:\\ 
\textbf{Clean Sample Accuracy (CSA):} For normal users, the CSA measures the proportion of clean test samples containing no trigger that is correctly predicted to their ground-truth classes. \\
\textbf{Attack Success Rate (ASR):} For an attacker, we represent the output of the backdoored model $M_{{\Theta^{adv}}}$ on poisoned input data $x^{poisoned}$ as $y^{'}= M_{{\Theta^{adv}}}(x^{poiosned})$ and the attacker's expected target as $t$. This index measures the ratio of $y^{'}$ which equals the attacker target $t$. This measurement also shows whether the neural network can identify the trigger pattern added to the input images.
\subsection*{DFT-based backdoor attack}
\begin{figure}
  \centering
  \includegraphics[width=0.9\linewidth]{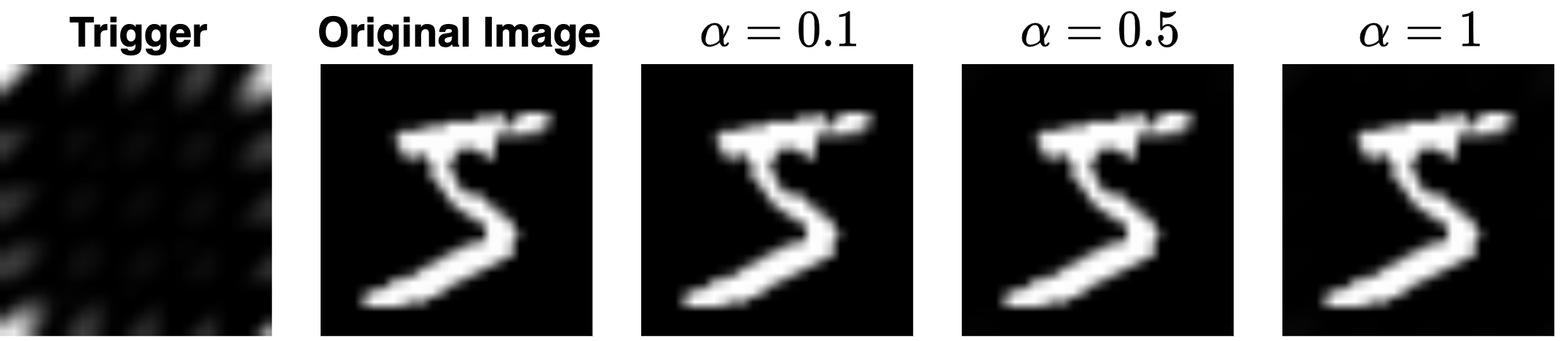}
  \caption{The figure shows the spatial trigger, original image and the poisoned samples generated by DFT-based method with $\alpha=0.1$, $\alpha=0.5$ and $\alpha=1$ respectively.}%
  \label{fig:mnist_five}
\end{figure}
In order to construct the poisoning training dataset with our DFT-based algorithm, we inject the frequency-domain trigger into 10$\%$ training data. For the images in which we plant the trigger, we replace their labels with our target label. In MNIST, CIFAR-10, and Imagenette, we select digit 5, "deer", and "building" as our targets respectively. We apply an energy factor $\alpha$ to control the invisibility of the poisoned images. 
To make the neural network learn the features of our frequency-domain trigger, we retrain the baseline models on the poisoning dataset with a small learning rate. When validating the backdoored model, we hide our trigger on the original validation dataset using the same $\alpha$ value, and then we compute their Clean Sample Accuracy (CSA) and Attack success Rate (ASR) (see \textbf{Section 4.1}). \par
\begin{table}
\centering
\begin{tabular}{cccc}
\toprule
  &$\alpha=0.1$ & $\alpha=0.5$  &  $\alpha=1$ \\
\midrule
Epoch & 48 & 10 & 3\\
CSA   & 98.53\%  & 98.31\% & 98.89\%   \\
ASR    & 98.48\%  & 99.99\% & 99.99\% \\
$l_{2}$ &0.0122 & 0.0610 &  0.1219\\
\bottomrule
\end{tabular}
\caption{DFT-based backdoor attack performance and $l_{2}$ values for different $\alpha$ values on MNIST.}
\label{tab:mnist_performance}
\end{table}
\paragraph{DFT-based method for gray images.}First, to demonstrate the feasibility of our attack, we conduct experiments on MNIST. Figure~\ref{fig:mnist_five} shows the poisoned samples generated on MNIST using different $\alpha$ values, and the first image shows the highlighted trigger pattern generated by our method for grayscale images. Table~\ref{tab:mnist_performance} shows the performance of our attack on MNIST using different $\alpha$ values. During the process of our experiments, we find that the smaller the $\alpha$ value, the slower the model converges and the more epochs are needed for training, which indicates that it is more difficult for our model to capture such slight perturbations. However, both ASR and CSA end up close to 100\%. Additionally, the $l_{2}$ value of the perturbation at $\alpha=0.1$ reaches only 0.0122 without affecting the performance, which means our model can detect the subtle change in image's frequency domain, thus making the trigger invisible.\par
\begin{figure}
  \centering
  \includegraphics[width=0.9\linewidth]{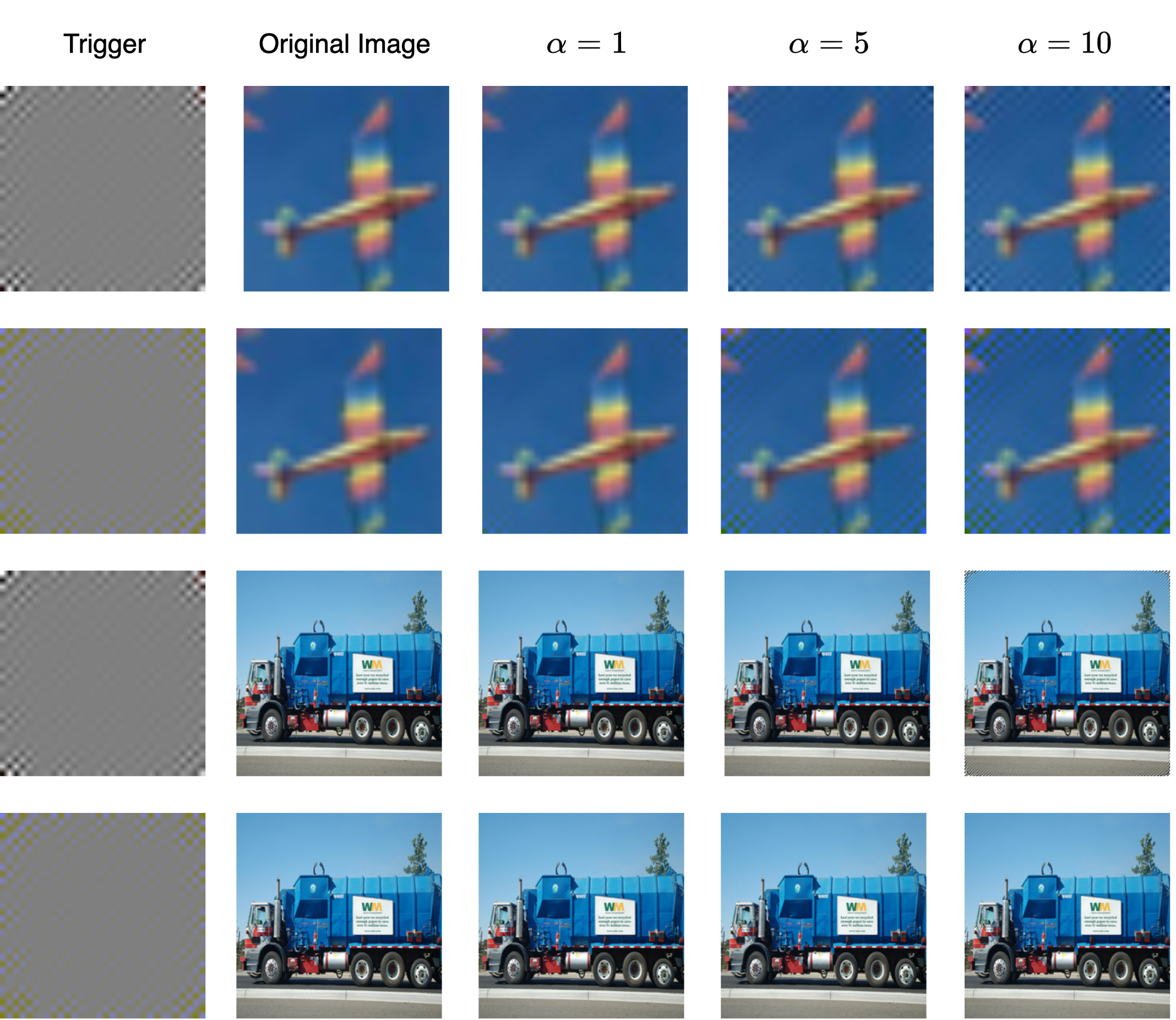}
  \caption{The figure shows trigger, original image and the poisoned samples generated by DFT-based method using different $\alpha$ values on CIFAR-10 (row ${1,2}$) and Imagenette (row ${3,4}$). Row ${1,3}$ and row ${2,4}$ respectively demonstrates two different triggers proposed in figure~\ref{fig:DFT_RGB}. }%
  \label{fig:multi_alpha}
\end{figure}
\begin{figure*}[tbp]
    \centering
    \includegraphics[width=2\linewidth]{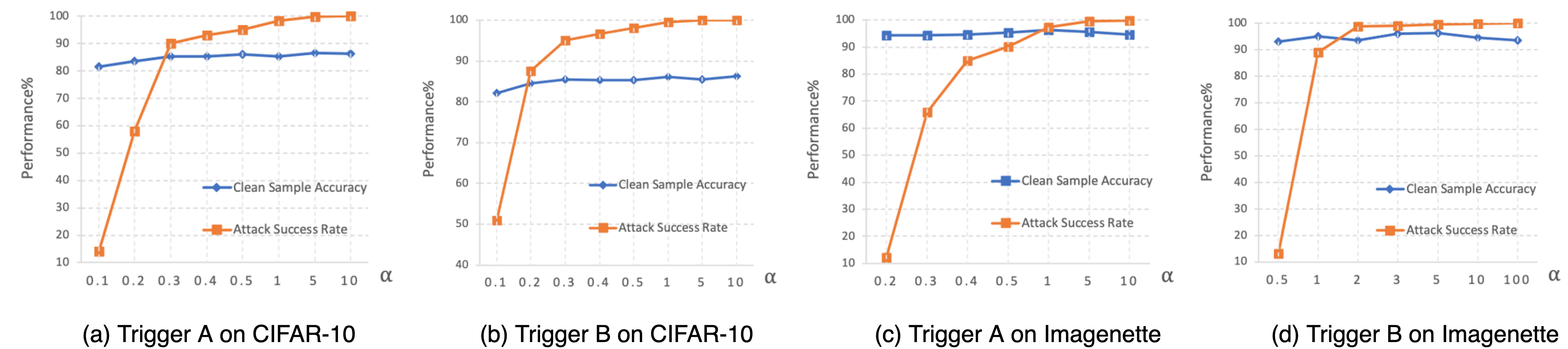}
    \caption{The relationship of the attack and invisibility with the $\alpha$ value increasing on CIFAR-10 and Imagenette datasets. Note that for each dataset, we implement two methods of generating poisoned samples for RGB images proposed in figure~\ref{fig:DFT_RGB}.}%
    \label{fig:four_alpha}  
\end{figure*}
\paragraph{DFT-based method for RGB images.}To evaluate the performance of the attack on the trigger strength of poisoned samples on CIFAR-10 and Imagenette, we carried out extensive experiments which are summarized in Figure~\ref{fig:four_alpha}. According to our two methods of crafting poisoned samples for RGB images proposed in Figure~\ref{fig:DFT_RGB}, we set different $\alpha$ values on CIFAR-10 and Imagenette to perform several backdoor attacks and test the attack success rate(ASR) as well as clean sample accuracy(CSA). Figure~\ref{fig:multi_alpha} shows the generated poisoned samples using different alpha values of two triggers. From four subfigures in figure~\ref{fig:four_alpha}, we can see that the ASR generally increases by boosting $\alpha$. Besides, in figure~\ref{fig:four_alpha}(a)(b), even when $\alpha=0.3$, the poisoned samples can be misclassified as our target with accuracy larger than 90.0$\%$. The effectiveness of conventional backdoor attack can be further enhanced by considering our method with $\alpha=0.3$ on CIFAR-10, which still guarantee the attack concealment without compromising the ASR and CSA. As for Imagenette, the best tradeoff point is $\alpha=0.5$ for Trigger A and $\alpha=1$ for Trigger B.\par
\begin{figure}
  \centering
  \includegraphics[width=\linewidth]{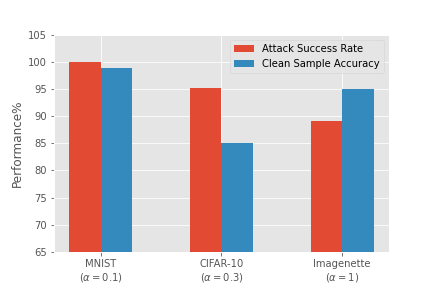}
  \caption{Attack performance on MNIST, CIFAR-10 (Trigger B), and Imagenette (Trigger B) with the best invisibility. Note that we choose the lowest $\alpha$ values while ensuring high ASR and CSA.}%
  \label{fig:performance_three_dataset}
\end{figure}

Besides, we perform experiments to compare the two DFT-based methods for RGB images on Imagenette, which are summarized in Table~\ref{tab:compare_triggerAB}. In the table, the "Best $\alpha$" indice indicates the lowest $\alpha$ values for Trigger A and Trigger B, respectively, while ensuring the ASR and CSA. Additionally, we apply $l_{2}$ paradigm and LPIPS to evaluate the invisibly of the two methods. From the table, we find that in contrast to Trigger A, Trigger B has better invisibility without sacrificing ASR and CSA.
\begin{table}
\centering
\begin{tabular}{cccccc}
\toprule
 Trigger & Best $\alpha$ & ASR  & CSA & $l_{2}$ & LPIPS \\
\midrule
A    &0.5 & 90.14\% & 95.26\% &1.057 &1.9e-3 \\
B    &1 & 89.11\% & 95.06\% &0.914 &7.8e-4\\
\bottomrule
\end{tabular}
\caption{Comparison between Trigger A and Trigger B proposed in figure~\ref{fig:DFT_RGB}. }
\label{tab:compare_triggerAB}
\end{table}
Figure~\ref{fig:performance_three_dataset} illustrates the accuracy of the backdoored model on the clean images (CSA) and the validation poisoning dataset (ASR). From the figure, it is clear that we can stealthily achieve our DFT-based backdoor attack on MNIST ($\alpha=0.1$), CIFAR-10 ($\alpha=0.3$) and Imagenette ($\alpha=1$) while hiding the trigger from human perception, indicating that our backdoored model can accurately identify the subtle changes in the image's frequency domain and simultaneously achieve the misclassification of the network.\par

\begin{figure}
  \centering
  \includegraphics[width=0.86\linewidth]{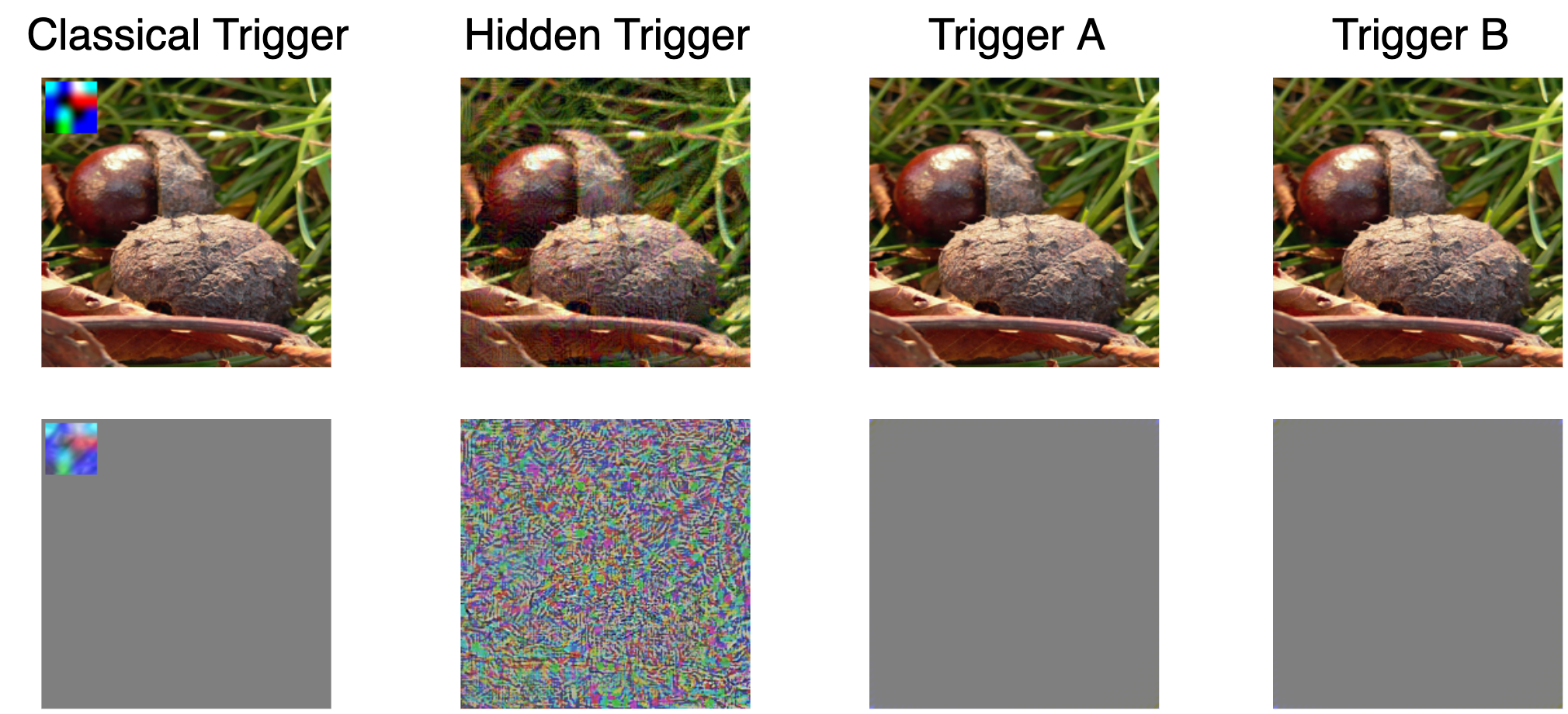}
  \caption{This figure shows poisoned samples generated by four methods and their corresponding trigger. Column 1,2 demonstrate the classical method and the hidden trigger backdoor, respectively. Column 3,4 illustrate our two methods for RGB images, respectively.}%
  \label{fig:comparison_four_mehtods}
\end{figure}

\subsection*{Comparison with classical attack}
We also conducted several experiments to compare our two methods with classical backdoor attack. Figure~\ref{fig:comparison_four_mehtods} shows different backdoored samples and their corresponding triggers. 
\begin{table}
\centering
\begin{tabular}{ccccc}
\toprule
 &\tabincell{c}{Classical\\Trigger} 
 &\tabincell{c}{Trigger A\\$\alpha=0.5$} 
 &\tabincell{c}{Trigger B\\$\alpha=1$}\\
\midrule
$l_2$ &  36.40 & \textbf{1.057} &\textbf{0.914} \\
LPIPS   & 0.049  & \textbf{1.9e-3} & \textbf{7.8e-4} \\
\bottomrule
\end{tabular}
\caption{Comparison with other works.}
\label{tab:comparison}
\end{table}
To prove the stealthiness of our method, we compute $l_2$ values and LPIPS indices of the four types of triggers used in classical backdoor~\cite{liu2017trojaning} and two DFT-based method for RGB images proposed in figure~\ref{fig:DFT_RGB}. For our two methods, we select $\alpha$ values used in table~\ref{tab:compare_triggerAB}.\par
$l_2$ value is used to calculate the euclidean distance between the backdoored image and the original image, so a lower value indicates the images are more similar. Recall that the LPIPS score measures the perceptual distance between the reference image and the blurred image. The range of LPIPS score is $[0,1)$. If two images are identical, the value is 0. A lower LPIPS value means two images are more similar; a higher score means the images are more different. A comparison of the $l_2$ paradigm value and LPIPS score for each attack is illustrated in Table~\ref{tab:comparison}. Our method achieves lower $l_2$ value and LPIPS (near 0). This demonstrates that it is more difficult for humans to distinguish our poisoned images from original images.\par

\section*{Availability of data and materials}
The dataset analysed during the current study was taken from  https://github.com/VinAIResearch.
\subsection*{Funding}
\subsection*{Acknowledgements}

\section*{Conclusion}
In this paper, we propose a novel method to add the backdoor trigger in the frequency domain of original images to generate poisoned samples. The poisoned data looks similar to the original images and does not reveal the trigger. Therefore, it is invisible enough to evade the event of a human visual inspection. Experiments on three different datasets demonstrate that our method implements invisible backdoor attacks without compromising the ASR and CSA. Additionally, we use two image similarity evaluation metrics to compare our method with a conventional backdoor attack and a state-of-the-art hidden trigger backdoor attack. We find that our approach adds the smallest perturbation. We believe such invisible backdoor attacks reveal the vulnerabilities of deep neural networks that need to be deployed in critical real-world applications.\par

\bibliographystyle{bmc-mathphys} 
\bibliography{bmc_article}     
\end{document}